\documentclass[a4paper,11pt]{article}
\pdfoutput=1 % if your are submitting a pdflatex (i.e. if you have
\usepackage{jcappub} % for details on the use of the package, please
\usepackage[T1]{fontenc} % if needed
\usepackage{xcolor}
\usepackage{amsmath, amsfonts}
\usepackage{tikz}
\usetikzlibrary{decorations.pathmorphing} % For wavy lines
\usepackage{booktabs}
\usepackage{amssymb}
\usepackage{physics}
\usepackage{mathtools}
\usepackage{multirow}
\usepackage{dsfont}
\usepackage{subfigure}
\usepackage[T1]{fontenc}
\usepackage{pdfpages}
\usepackage{mathrsfs}
\usepackage{amsmath,amsfonts, times}
\usepackage{bm} %include bold math: \bm{} creates bold letters in math mode
\usepackage[normalem]{ulem}
\usepackage{epstopdf}
\begin{document}
\title{\boldmath Braneworlds in Einstein-Scalar-Gauss-Bonnet gravity}

\author[a]{José Euclides G. Silva}
\author[b]{Leandro A. Lessa}
\author[a]{Roberto V. Maluf}
\affiliation[a]{Universidade Federal do Cear\'a (UFC), Departamento de F\'isica,\\ Campus do Pici, Fortaleza - CE, C.P. 6030, 60455-760 - Brazil.}
\affiliation[b] {Programa de Pós-graduação em Física, Universidade Federal do Maranhão, Campus Universitário do Bacanga, São Luís (MA), 65080-805, Brazil.}
\emailAdd{euclides@fisica.ufc.br}
\emailAdd{leandrophys@gmail.com}
\emailAdd{r.v.maluf@fisica.ufc.br}

\abstract{
We explore the features of a thick braneworld model in five dimensions governed by a Einstein-Gauss-Bonnet gravity with a non-minimal coupling to a dynamical scalar field. We consider two possible scalar-GB coupling function $\chi(\phi)$, one parity-even and another parity-odd function of the scalar field $\phi$. For both choices, the scalar-Gauss-Bonnet non-minimal coupling produces a warped asymptotically $AdS_5$ spacetime even in the absence of a bulk cosmological constant. Outside the brane core, a negative cosmological constant is bounded by the scalar-GB coupling function. For a thick 3-brane configuration, we found solutions with localized brane energy density and pressure that dynamically produce a bulk cosmological constant. The corresponding scalar field solutions exhibit a non-topological (domain wall) behavior. In order to probe the 3-brane stability solution, we employed a perturbative analysis, by perturbing the thick brane solutions up to first-order. The Kaluza-Klein (KK) tensorial gravitational modes possess a localized massless mode and a tower of non-tachyonic diverging massive modes, what renders the solutions stable at least at the perturbative level.} 
\maketitle

\flushbottom
%%%%%%%%
%\date{\today}
%%%%%%%%
%%%%%%%%%%%%%%%%%%%%%%%%%%%%%%%%%%%%%%%%%%%%%%

\section{Introduction}

\indent\indent 
General Relativity (GR) stands as the most successful gravitational theory to date. Alongside Quantum Mechanics, it forms one of the foundational pillars of modern physics. Einstein’s theory has undergone exhaustive experimental scrutiny, and no deviations from its predictions have been detected so far \cite{Will,Berti}. However, the non-renormability of the Einstein gravity \cite{tHooft,Goroff} and the current cosmological acceleration \cite{acc} suggest that GR might be an effective theory of a quantum theory of gravity such as string theory \cite{Gross1} or loop quantum gravity \cite{lqg}. Departures from the GR can be probed by
adding higher-order curvature terms into the Einstein-Hilbert (EH) action \cite{stelle}. These terms can be obtained by quantum correction on the EH action. For instance, 1-loop corrections of the EH action lead to inclusion of quadratic curvature terms (e.g., $R^2$, $R^{\mu\nu}R_{\mu\nu}$ and $R^{\mu\nu\alpha\beta}R_{\mu\nu\alpha\beta}$)  \cite{stelle}, albeit it might introduce ghost modes or unitarity violations \cite{Sotiriou,Salvio}.

A clever way to circumvent these pathologies is to construct effective gravitational theories with higher-curvature terms that do not introduce higher-derivative terms in the equations of motion (EOM). The most general d-dimensional gravitational theory that produces second-order EOM for any spacetime is Lovelock gravity \cite{Lovelock:1971yv,Lovelock:1972vz}.
%This theory is also supported by a uniqueness theorem, which states—under certain non-trivial conditions (for details, see Refs.\cite{Berti:2015itd,Sotiriou:2015pka} )—that GR (described by the Einstein-Hilbert term plus a cosmological constant) is the only diffeomorphism-invariant gravitational theory in $d=4$ that produces second-order EOM.
The quadratic curvature term in Lovelock theory is given by the well-known Gauss-Bonnet term , i.e., $\mathcal{G}= R^2-4R^{\mu\nu}R_{\mu\nu}+R^{\mu\nu\alpha\beta}R_{\mu\nu\alpha\beta} $, which becomes topological in $d=4$ but introduces new physical effects in higher dimensions \cite{pad}. 
%Numerous studies have explored this quadratic term [\textcolor{red}{CITAR ARTIGOS}].
However, it is possible to bypass Lovelock theorem within the framework of effective field theory by relaxing some of its assumptions. One way to circumvent the Lovelock theorem is achieved by considering a non-minimal coupling between the gravitational field and a real scalar field $\phi$ of form $\chi(\phi)\mathcal{G}$ \cite{Mignemi,Sotiriou1} . This theory, known as Einstein-scalar-Gauss-Bonnet (ESGB) gravity, naturally arises in the low-energy limit of string theory \cite{Callan,Kanti}.  In $d=4$ this coupling breaks the topological nature of the GB term, leading to interesting features, such as hairy black holes \cite{hair1,hair2} and spontaneous black hole scalarization \cite{segb1,segb2,segb3}.

In this work, we consider the effects of a Einstein-Scalar-Gauss-Bonnet (ESGB) theory with one spacelike extra dimension. In particular, we adopt the braneworld approach, which considers our $3+1$ universe as a hypersurface ($3-$brane) embedded in a $d=5$ curved spacetime. These scenarios, initially proposed to solve the gauge hierarchy problem in a $AdS_5$ bulk spacetime \cite{Randall}, also provide geometric solutions for the dark energy and dark matter problems. By adding a self-interacting scalar field minimally coupled to the Einstein gravity, the resulting scalar-Einstein theory produces stable domain-wall thick 3-branes. The scalar potential vacuum not only leads to a topological stable configuration but also dynamically produces the bulk cosmological constant. Extended braneworld models were proposed considering codimension two dimensions \cite{conifold,liucod2}, Weyl geometry \cite{weyl}, Rastall gravity \cite{rastal}, torsion \cite{torsion1,torsion2} and others. 

Braneworld models with modified gravitational dynamics have been studied in the literature, such as in $f(R)$ \cite{ft} and in the teleparallel $f(T)$ \cite{ft}, $f(Q)$ \cite{fq} gravities. The effects of the Gauss-Bonnet term on branes dynamics was studied in \cite{branegb1,branegb2}, brane cosmology in \cite{gbbranecosmology}, $f(\mathcal{G})$ in \cite{branefgb} and non-minimal couplings \cite{gbnonmin1,gbnonmin2,gbnonmin3}. Despite having quadratic term of the curvature in the action, the additional terms in the EoM are up to second-order in the metric (warp factor) \cite{gb}. For a minimally coupled scalar field, the well-known first-order formalism \cite{fo} can be applied, furnishing topological new stable thick brane solutions \cite{gb}.

By allowing a non-minimal coupling $\chi(\phi)\mathcal{G}$, we found non-trivial modifications of the scalar field and the geometry. Although the EOM contains terms up to second derivatives, new terms involving $\chi$, $\chi_\phi$ and $\chi_{\phi\phi}$ arise, modifying the scalar field vacuum and adding new non-trivial terms as $\chi_{\phi\phi}A'^2 \phi'^2$ into the gravitational equations. As a result, an upper bound for the bulk cosmological constant is found such that the spacetime is asymptotically $AdS_5$. Moreover, the non-minimal coupling allows an asymptotic $AdS_5$ geometry even in the absence of a cosmological constant. Similar features were also found in another high curvature model, the so-called warped Einsteinian cubic gravity \cite{cubic}. For a thick brane solution, despite having a localized source (energy and pressure densities), the scalar field does not exhibit a domain-wall profile, but a non-topological behavior similar to one found in \cite{nontop}. The linear perturbation of the gravitational and source equations reveals that the tensor modes gravitational Kaluza-Klein (KK) modes depends not only on the bulk geometry but on the non-minimal coupling as well. Yet, for the linear and quadratic non-minimal coupling considered, the KK spectrum has no tachyonic state and allows a localized massless mode.
%For a thick 3-brane solution sourced by a non-minimally coupled scalar field, the ESGB term introduces non-trivial effects on the model's dynamics and stability, with crucial dependence on the coupling function form. Beyond analyzing the warped gravitational solutions in this scenario, we investigate the propagation of tensor gravitational modes in these modified brane-world geometries. We find that the perturbation equation for Kaluza-Klein (KK) gravitational modes acquires explicit dependence on the scalar field dynamics. In the simplified case where scalar field perturbations are neglected, we demonstrate that no tachyonic KK gravitational modes appear, indicating linear stability of the tensor sector in these configurations.

The work is organized as follows. In  section \eqref{sec2}, we construct the ESGB gravity in five dimensional warped geometry, imposing some shape for scalar-curvature coupling.  In  section~\eqref{sec3}, we investigate the effects of the scalar field non-minimally coupled to the Gauss-Bonnet term on the dynamics of a thin brane. Subsequently, in  section~\eqref{sec4}., we analyze the modifications induced by this theory in the thick brane scenario
%present the equations of motion of cubic theory for a warped metric and analyze the modifications in the thin bran solution due to quartic and cubic interaction.
%In section \ref{sec3} we specialize the analysis for a five dimensional warped geometry, imposing a condition on the cubic coefficients in order to the gravitational equations to have terms up to second-order. Then, we study the effects of the quadratic and cubic term on thin 3-brane solutions in subsection \ref{sec3.1} and for thick brane solutions generated by scalar field using the first order formalism in the subsection \ref{sec3.2}. In section \ref{sec4}, we obtain the linearized equations for gravitational fluctuations up to cubic order explicitly.
In section \eqref{sec5}, we analyze the stability of the model and the problem of Kaluza-Klein (KK) mode localization. Final remarks are summarized in section \eqref{con}. Throughout the text, we adopt the capital Roman indices ($A,B,... = 0,1,2,3,4$) denote 5-dimensional bulk spacetime indices, the Greek indices ($\mu , \nu , ... = 0,1,2,3$) the spacetime indices of the braneworld.
%and small Roman indices ($a, b, ... = 0,1,2,3$) the coordinetes of the brane. 
Moreover, we adopt the metric signature $(-,+,+,+)$.

\section{Warped spacetime in Einstein-Scalar-Gauss-Bonnet gravity} \label{sec2}

In this section, we introduce a five-dimensional braneworld model that extends the usual Einstein-Hilbert term by incorporating a scalar field $\phi$ that couples non-minimally to gravity through a general coupling function $\chi(\phi)$ associated with the quadratic Gauss-Bonnet term. Such a theory is described by the following action
\begin{align}
\label{sgbaction}
S & =\int d^{5}x\sqrt{-g_{5}}\left\{ \frac{R}{2\kappa_{5}}-\frac{1}{2}\mathcal{D}_{A}\phi\mathcal{D}^{A}\phi-V(\phi)+\alpha\chi(\phi)\mathcal{G}\right\}, 
\end{align}
where $\kappa_5 = \frac{8\pi}{M_{Pl}^3}$ is the
Planck mass scale in $d=5$ and  $V(\phi)$ is the potential associated to the scalar field, $\alpha$ is a constant free parameter controlling the contribution of the Gauss-Bonnet term defined as
\begin{equation}
    \mathcal{G}=R^{2}-4R_{AB}R^{AB}+R_{ABMN}R^{ABMN}.
\end{equation}
The scalar field $\phi$ is coupled to $\mathcal{G}$, which has mass dimension $4$, through a function $\alpha\chi(\phi)$, with mass dimension $[\alpha]=1$. For $\chi(\phi)=1$, we recover the usual Einstein-Gauss-Bonnet action with a bulk scalar field \cite{gb}. Thus, the non-minimal coupling function $\chi(\phi)$ turns the Gauss-Bonnet coupling constant $\alpha$ into a position-dependent coupling. Moreover, different choices of the function $\chi(\phi)$ correspond to
different ESGB gravity theories.

By varying the action in Eq. (\ref{sgbaction}) with respect to the metric $g^{AB}$ we obtain the modified gravitational equations
\begin{equation}\label{EoMmetric}
    G_{AB}+\kappa\left[g^{(5)}{}_{AB}\bigl(V(\phi)+\tfrac{1}{2}\mathcal{D}_{N}\phi\mathcal{D}^{N}\phi\bigr)-\mathcal{D}_{A}\phi\mathcal{D}_{B}\phi\right]+\kappa\alpha\mathcal{J}_{AB}=0,
\end{equation}where $\mathcal{D}_{A}$ denotes the covariant derivatives, $G_{AB}\equiv R_{AB}-\tfrac{1}{2}g^{(5)}{}_{AB}R$ is the Einstein's tensor, and $\mathcal{J}_{AB}$ is a symmetric 2-tensor, defined in terms of the variation of the the Scalar-Gauss-Bonnet term as
\begin{align}
\mathcal{J}_{AB}\equiv\chi\left(4R_{AB}R-8R_{A}{}^{L}R_{BL}-8R^{LM}R_{ALBM}+4R_{A}{}^{LMP}R_{BLMP}\right)\nonumber\\
+\chi_{\phi}\left(-4R\mathcal{D}_{A}\mathcal{D}_{B}\phi-8R_{AB}\mathcal{D}_{Q}\mathcal{D}^{Q}\phi+8R_{AQ}\mathcal{D}^{Q}\mathcal{D}_{B}\phi\right.\nonumber\\
\left.+8R_{BQ}\mathcal{D}^{Q}\mathcal{D}_{A}\phi+4R_{AQBR}\mathcal{D}^{R}\mathcal{D}^{Q}\phi+4R_{ARBQ}\mathcal{D}^{R}\mathcal{D}^{Q}\phi\right)\nonumber\\
+\chi_{\phi\phi}\left(8R_{AT}\mathcal{D}_{B}\phi\mathcal{D}^{T}\phi+8R_{BT}\mathcal{D}_{A}\phi\mathcal{D}^{T}\phi-4R\mathcal{D}_{A}\phi\mathcal{D}_{B}\phi\right.\nonumber\\
\left.-8R_{AB}\mathcal{D}_{T}\phi\mathcal{D}^{T}\phi+8R_{ATBZ}\mathcal{D}^{T}\phi\mathcal{D}^{Z}\phi\right)\nonumber\\
+g^{(5)}{}_{AB}\left[\chi\left(-R_{PQMN}R^{PQMN}+4R^{QM}R_{QM}-R^{2}\right)\right.\nonumber\\
+\chi_{\phi}\left(4R\mathcal{D}_{M}\mathcal{D}^{M}\phi-8R_{MN}\mathcal{D}^{M}\mathcal{D}^{N}\phi\right)\nonumber\\
\left.+\chi_{\phi\phi}\left(4R\mathcal{D}_{G}\phi\mathcal{D}^{G}\phi-8R_{GH}\mathcal{D}^{G}\phi\mathcal{D}^{H}\phi\right)\right],
\end{align} where $\ensuremath{\chi_{\phi}=\frac{d\chi}{d\phi}}$ and $\ensuremath{\chi_{\phi\phi}=\frac{d^{2}\chi}{d\phi^{2}}}$.
 The EOM for the scalar field $\phi$ has the form
\begin{equation}
\label{scalareom}
g^{(5)}{}_{AB}\mathcal{D}^{A}\mathcal{D}^{B}\phi-V_{\phi}(\phi)+\alpha\chi_{\phi}\left(R^{2}-4R_{AB}R^{AB}+R_{ABMN}R^{ABMN}\right)=0.
\end{equation}
Note that, for $\chi=1$ the Eq.(\ref{EoMmetric}) and Eq.(\ref{scalareom}) become the usual Einstein-GB equations \cite{gb}.

The scalar field EOM in Eq.(\ref{scalareom}) is a self-interacting non-linear equation, with a non-minimal coupling function $\chi(\phi)$.
In this work, we consider two possible choices for the non-minimal coupling function $\chi(\phi)$, namely the exponential or dilaton-like coupling~\cite{Mignemi,Kanti}
\begin{equation}
\label{dilatonchi}
    \chi(\phi)=e^{\lambda\phi},
\end{equation}
and the quadratic coupling
\begin{equation}
\label{pdm}
    \chi(\phi)=-\frac{M^2}{2} \phi^2,
\end{equation}
%and the quartic coupling
%\begin{equation}
%\label{quariccoupling}
%    \chi(\phi)=\frac{\gamma}{4}(\phi^2 - a^2)^2,
%\end{equation}
where $\lambda$ and $M$ are real constants. Note that in quadratic ESGB gravity, the theory exhibits $\phi \rightarrow -\phi$ symmetry. Moreover, this choice is widely studied in the context of spontaneous scalarization of black holes~\cite{segb1,segb2}, as it represents the simplest case where a tachyonic instability must be present, and the dominant-order term is expected to control the onset of the instability. For $\lambda\rightarrow 0$, the dilaton-like coupling can be approximated as
\begin{equation}
\label{dilatonchiapprox}
    \chi(\phi)=1+\lambda\phi +\mathcal{O}(\lambda^2).
\end{equation}
Thus, we approximately recover shift-symmetric ESGB gravity~\cite{segb1,segb2,segb3}, meaning the $\phi \rightarrow -\phi$ symmetry is no longer valid (considering only up to the leading-order linear term).

Let us consider a warped braneworld in the form \cite{gb}
\begin{equation}
\label{rsmetric}
ds^{2}_{5}=e^{2A(y)}\eta_{\mu\nu}dx^{\mu}dx^{\nu}+dy^2,
\end{equation}
where the function $A(y)$ is the so-called warp factor. In addition, let us consider an extra-dimension dependent scalar field $\phi=\phi(y)$. Accordingly, the gravitational equations~\eqref{EoMmetric} and the scalar field equation (\ref{scalareom}) yields to

\begin{align}
\label{einstein00}
6(A''+2A'^{2})+\kappa\bigl(\phi'^{2}+2V\bigr)\nonumber \\
-48\alpha\kappa\left[\chi A'^{2}\left(A''+A'^{2}\right)+\chi_{\phi} A'\left(2\phi'A''+A'(\phi''+3\phi'A')\right)+\chi_{\phi\phi}A'^{2}\phi'^{2}\right] & =0,
\end{align}

\begin{equation}
\label{einstein44}
12A'^{2}-\kappa\bigl(\phi'^{2}-2V\bigr)-48\alpha\kappa\left(\chi A'^{4}+4\chi_{\phi}A'^{3}\phi'\right)=0,
\end{equation}
\begin{equation}
\label{phieom}
\phi''+4\phi'A'-V_{\phi} +24\alpha\chi_{\phi} A'^{2}(4A''+5A'^{2})=0.
\end{equation}
The prime stands for derivative with respect to $y$. We observe that the modified gravitational field equations presented above contain no field derivatives beyond second order.
Moreover, the non-minimal scalar-GB coupling yields new terms depending on $\chi$, $\chi_\phi$ and $\chi_{\phi\phi}$. For $\chi=1$, the usual Einstein-GB braneworld equations are recovered \cite{gb}.
%Furthermore, we note that the scalar field coupling and Gauss-Bonnet term exhibit non-trivial mixing. In contrast to Ref. [chineses], the present model features strong curvature-scalar coupling. In the following section, we examine the equation system~\eqref{einstein00},~\eqref{einstein44} and~\eqref{phieom}  under different dynamical regimes and coupling configuration $\chi(\phi)$.}

%%%%%%%%%%%%%%%%%%%%%%%%%%%%%%%%%%%%%%%%%%%%%%%%%%%%%%%%%%%%%%%%%%%%%%%%%%%%%%%%%%%%%%%%%%%%%%%%%%%%%%%%%%%%%%%%%%%%%  END OF SECTION 2 %%%%%%%%%%%%%%%%%%%%%%%%%%%%%%%%%%%%%%%%%%%%%%%%%%%%%%%%%%%%%%%%%%%%%%%%%%%%%%%%%%%%%%%%%%%%

\section{Outside the brane core} \label{sec3}

In the region exterior of the brane core, i.e., for $y\rightarrow \infty$, let us assume that the scalar field reaches a constant $\phi_0 \neq 0$ such that $\phi'_0 =0$. Moreover, let us assume that the potential also has a non-vanishing value, i.e., $V(\phi_0)\neq 0$. This non-vanishing value of $V$ enables us to dynamically define the bulk cosmological constant $\Lambda=2\kappa V(\phi_0)$. 

Assuming a Randall-Sundrum (RS) ansatz for the warp factor of form 
$A'(y)=-c$ , the gravitational equations ~\eqref{einstein00} and~\eqref{einstein44}  assume the form
\begin{equation}
\label{rsequation}
    12c^2 + \Lambda - 48\alpha\kappa \chi(\phi_0)c^4 = 0.
\end{equation}
For $\alpha =0$, the we obtain the usual RS solution $c^{2}=-\Lambda$, i.e., the exterior region is a $AdS_5$ spacetime. On the other hand, for $\alpha\neq 0$ we obtain
\begin{equation}
    c^2 = \frac{1}{8\kappa \alpha\chi(\phi_0)}\left(1+\frac{\sqrt{3(3+4\kappa\alpha\chi(\phi_0)\Lambda)}}{3}\right).
\end{equation}
Note that the scalar field asymptotic value $\phi_0$ not only defines the bulk cosmological constant but also controls the value of the GB constant $\alpha$.
For the dilaton coupling $\chi= 1+\lambda\phi_0$, we obtain the relation
\begin{equation}
    c^2 = \frac{1}{8\kappa \alpha(1+\lambda\phi_0)}\left(1+\frac{\sqrt{3(3+4\kappa\alpha(1+\lambda\phi_0)\Lambda)}}{3}\right)
\end{equation}
A noteworthy feature is that, for $\Lambda<0$, the cosmological constant  must satisfy the constrain
\begin{equation}
\label{rssolution}
    |\Lambda|\leq \frac{3}{4\alpha \kappa(1+\lambda\phi_0)}.
\end{equation}
Thus, a $AdS_5$ asymptotic bulk is allowed as long as the condition (\ref{rssolution}) is satisfied. For the quadratic coupling $\chi(\phi)=-\frac{M^2}{2}\phi^2$, we obtain
\begin{equation}
    c^2 = \frac{1}{8\kappa \alpha(1+\lambda\phi_0)}\left(1+\frac{\sqrt{3(3-2\kappa\alpha M^2 \phi_{0}^2\Lambda)}}{3}\right).
\end{equation}
Then, a $\Lambda<0$ solution is possible for any $\Lambda$.
%Finally, let us consider the quartic coupling in Eq.(\ref{quariccoupling}). Note that for $a=\phi_0$, then
%the Eq.(\ref{rsequation}) assumes the usual form $12c^2-\Lambda =0$. Therefore, if $a=\phi_0$ there is no changes on the exterior solution driven by the scalar-GB term.

Another noteworthy feature of the SGB term is that it leads to a non-trivial solution of Eq.(\ref{rsequation}) even for $\Lambda=0$. Indeed, for $\Lambda=0$, we obtain
\begin{equation}
    c=\frac{1}{2\sqrt{\kappa\alpha\chi(\phi_{0})}}.
\end{equation}
Therefore, the Gauss-Bonnet term allows a RS-like solution even without a Bulk cosmological constant. For $\chi(\phi)\approx (1+\lambda\phi)$, the exponential factor $c$ is given by
\begin{eqnarray}
    c &=&\pm \frac{1}{4\kappa\alpha (1+\lambda\phi_0)}\\
      &\approx & \pm \frac{1}{4\alpha\kappa} \mp \frac{\lambda\phi_0}{4\kappa\alpha}\nonumber.    
\end{eqnarray}
Note that for the usual Gauss-Bonnet term, i.e., for $\chi(\phi)=1$, already yields to an exponentially decreasing solution even in the absence of a bulk cosmological constant. Moreover, for the
quadratic coupling, we obtain
\begin{equation}
    c=\mp \frac{1}{2\kappa\alpha M^2 \phi_{0}^2}.
\end{equation}
Thus, the non-minimal coupling leads to an effective Gauss-Bonnet coupling $\alpha_{eff}=\alpha \chi(\phi_0)$ due to the non-minimal interacting function $\chi_{\phi}$.

Now let us turn out attention to the scalar field EOM in Eq.(\ref{phieom}). Assuming that $\phi\rightarrow \phi_0$ such that $\phi'_0 = \phi''_0 =0$, the scalar field EOM outside the brane core leads to 
\begin{equation}
    V_\phi \rightarrow 24\alpha\chi_{\phi}(\phi_0)c^2.
\end{equation}
For the usual Einstein gravity and Einstein-Gauss-Bonnet models, $\chi_\phi =0$, and hence $V_\phi \rightarrow 0$. This means that, in these models, the $\phi_0$ is a vacuum configuration. Thus, the SGB term shifts the field from the vacuum value by an amount proportional to $\chi_\phi (\phi_0)$. Note that, for a power-like coupling as $\chi(\phi)\approx (\phi^2 -\phi_{0}^2)^n$, the SGB term preserves the 
vacuum of the scalar field potential.

%%%%%%%%%%%%%%%%%%%%%%%%%%%%%%%%%%%%%%%%%%%%%%%%%%%%%%%5555

\section{Thick brane} \label{sec4}

Let us now consider the brane core effects. Subtracting the gravitational equations (\ref{einstein00}) and (\ref{einstein44}) yields to
\begin{equation}
\label{thickphiequation}
    (1-24\alpha\chi_{\phi\phi}A'^2)\phi'^2 -24\alpha\chi_\phi A'^2 \phi'' -24\alpha\chi_\phi A'(A''-A'^2)\phi' + \left(\frac{3}{\kappa} -24\alpha\chi A'^2\right)A''=0.
\end{equation}
The Eq.(\ref{thickphiequation}) for the scalar field has rather invoked terms involving $A'$, $A''$, $\phi'$ and $\phi''$. Moreover, the Eq.(\ref{thickphiequation}) strongly depends on the non-minimal coupling function $\chi(\phi)$ adopted.
Note that for $\alpha=0$ we obtain the usual braneworld equation
$\phi'^2 + \frac{3}{\kappa}A''=0$, which contains only $\phi'$ terms and it allows us to adopt the first-order formalism \cite{fo}. 
For the usual Gauss-Bonnet term, i.e., for $\chi(\phi)=1$, the Eq.(\ref{thickphiequation}) simplifies to
\begin{equation}
\label{gbeq}
    \phi'^{2}+\left(\frac{3}{\kappa} -24\alpha A'^2\right)A''=0,
\end{equation}
which contain only the first-derivative of the scalar field and the derivatives of the warp factor. Thus, a first-order formalism can be applied to this case, as performed in the Ref.\cite{gb}.

It is worthwhile to mention that, in order to obtain a regular solution in the brane core which asymptotically tends to $A'=-c$, it is usually assumed that inside the brane core
\begin{eqnarray}
\label{corecondition}
    A''<0   &,&    y\rightarrow 0,
\end{eqnarray}
whereas outside the brane core
\begin{equation}
\label{vacuumcondition}
A'\rightarrow - c.    
\end{equation}
Assuming that $\alpha\geq 0$, the equation (\ref{gbeq}) satisfies the condition in Eq.(\ref{corecondition}) provided that
\begin{equation}
    \alpha \leq \frac{1}{8\kappa A'^4}.
\end{equation}
Therefore, there is an upper bound for the Gauss-Bonnet constant $\alpha$ in order to obtain an smooth brane core.

By adding the Eq.(\ref{einstein00}) and Eq.(\ref{einstein44}) we obtain the expression for the scalar field potential $V$ as
\begin{eqnarray}
\label{poteq}
    V &=&-\frac{6}{\kappa}A'^2 -\frac{3}{2\kappa}A''+12\alpha\Big(\chi A'^2 (A''+2A'^2)+\chi_{\phi} A'[2\phi'A'' + A'(\phi'' + 7A'\phi')]\nonumber\\
    &+& \chi_{\phi\phi} A'^{2} \phi'^{2}\Big).
\end{eqnarray}
Thus, if the scalar field $\phi$ and the warp factor $A$ are known, it is possible to study the behaviour of the potential $V$ along the extra dimension using Eq.(\ref{poteq}).  Indeed, for the usual braneworld dynamics, i.e., for $\alpha=0$, outside the brane core, we obtain
\begin{equation}
    V(y\rightarrow\pm\infty) \approx -\frac{6c}{k}.
\end{equation}
Therefore, the bulk cosmological constant is dynamically generated by the vacuum value of the scalar field potential, $V(\phi_0)= V(y\rightarrow\pm\infty)$. Note that the negative value of $V(\phi_0)$ leads to an asymptotic $AdS_5/Z_2$ spacetime.

\begin{figure}[!ht] 
\centering
\includegraphics[height=4.4cm]{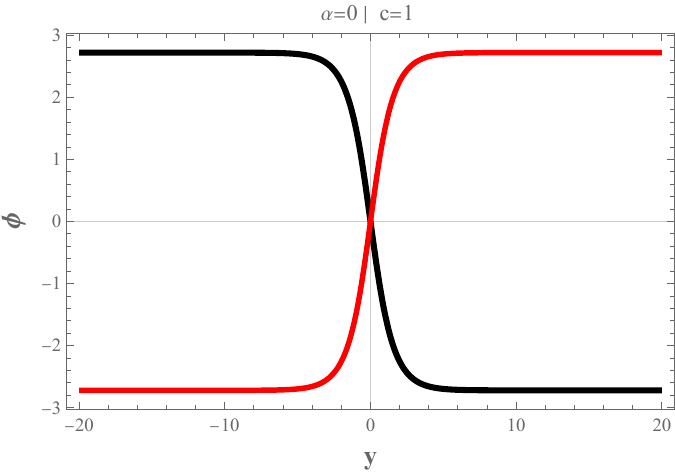}\quad
        \includegraphics[height=4.4cm]{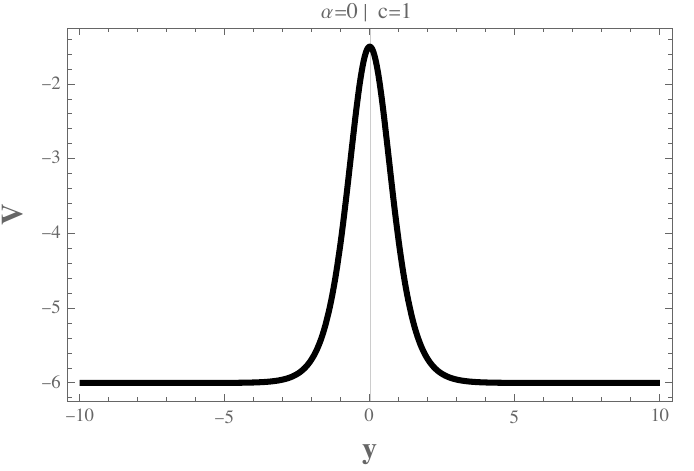}\quad
\caption{Scalar field profile and the potential with respect to $y$ for the usual sine-gordon brane in Einstein gravity.}  \label{usual}
\end{figure}

\begin{figure}[!ht] 
\centering
\includegraphics[height=4.4cm]{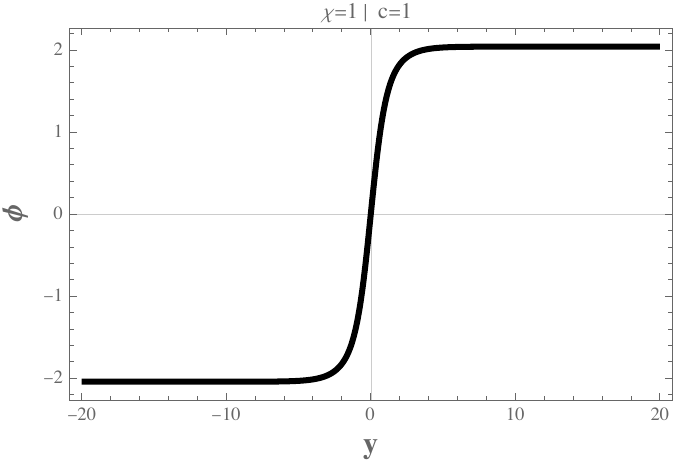}\quad
        \includegraphics[height=4.4cm]{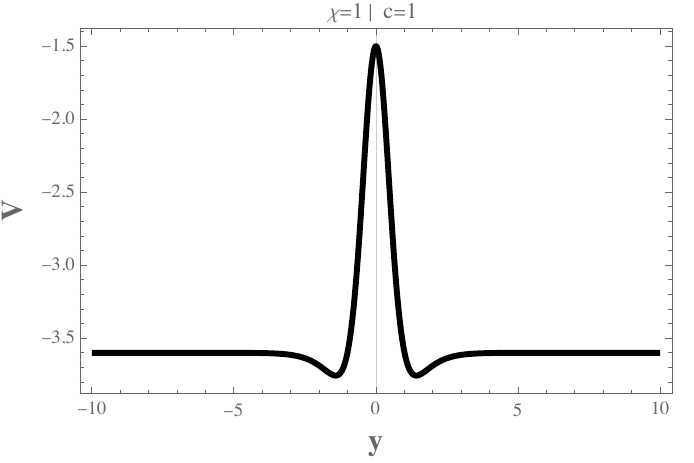}\quad
\caption{Scalar field profile and the potential with respect to $y$ for the Einstein-Gauss-Bonnet gravity, for $\alpha=0.1$.}  \label{gb1}
\end{figure}

Since the EOM are rather involved, the first-order formalism can not be applied. Instead, we assume a thick brane ansatz in the form
\begin{equation}
\label{warpansatz}
    A(y)=- \ln{\cosh{c y}}.
\end{equation}
The warp function in Eq.(\ref{warpansatz}) satisfies the core Eq.(\ref{corecondition}) and vacuum conditions in Eq.(\ref{vacuumcondition}). For $\alpha=0$, substituting the ansatz (\ref{warpansatz}) into the  Eq.(\ref{thickphiequation}), we obtain $\phi(y)=\frac{2\sqrt{3}}{\kappa}\arctan \tanh\left(\frac{cy}{2}\right)$, i.e., the sine-gordon domain wall solution, whose profile and the potential are depicted in the Fig.(\ref{usual}). Note that the asymptotic value of the potential generates the bulk cosmological constant. For the Einstein-Gauss-Bonnet case $(\chi=1)$, the Eq.(\ref{gbeq}) for the warp factor in Eq.(\ref{warpansatz}) also produces a domain-wall like configuration, as shown in Fig.(\ref{gb1}).

\begin{figure}[!ht] 
\centering
\includegraphics[height=4.4cm]{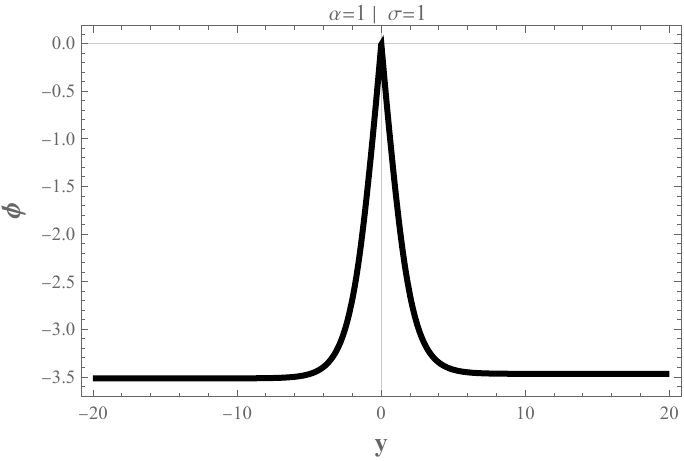}\quad
        \includegraphics[height=4.4cm]{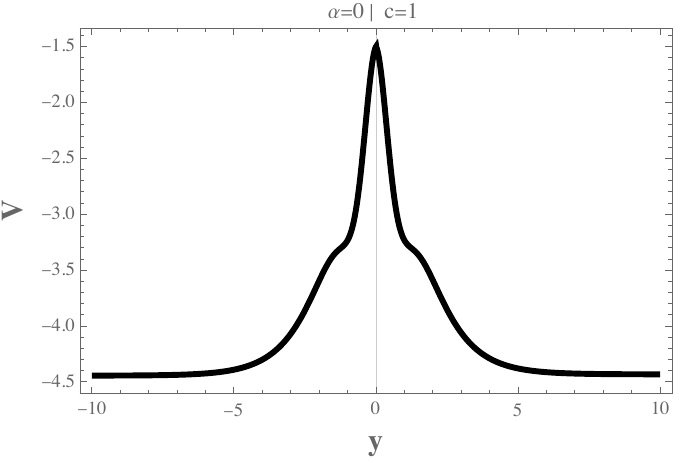}\quad
\caption{Scalar field profile and the potential with respect to $y$ for the scalar Gauss-Bonnet model coupling function $\chi=1+\lambda\phi$.}  \label{sgb1}
\end{figure}

However, the inclusion of the scalar coupling $\chi\neq0$ significantly alters the scalar field dynamics. Indeed, for the linear non-minimal coupling $\chi\approx 1+\lambda\phi$ coupling, the Eq.(\ref{thickphiequation}) reads
\begin{equation}
\label{philinear}
    \phi'^2 - 24\alpha\lambda A'^2 \phi'' -24\alpha\lambda A'(A''-A'^2)\phi' - 24\alpha\lambda A'^2 A'' \phi +\left(\frac{3}{\kappa}-24\alpha\lambda A'^2\right)A''=0
\end{equation}
Unlike in the usual Einstein gravity or in the Einstein-Gauss-Bonnet model, the Eq.(\ref{philinear}) contains terms proportional to $\phi''$, $\phi'$ and $\phi$. Thus, the Eq.(\ref{philinear}) does not allow us to adopt the first-order formalism, as performed for the usual EGB model \cite{gb}. However, we can solve the Eq.(\ref{philinear}) for the thick 3-brane anstaz in Eq.(\ref{warpansatz}).
The solution of Eq.(\ref{philinear}) for the ansatz in Eq.(\ref{warpansatz}) is shown in the Fig.(\ref{sgb1}). Note that the scalar field does not exhibit domain-wall profile. Thus, the
new terms steaming from the scalar coupling break the topological symmetry of the scalar field.
Nevertheless, it is worthwhile to mention that the potential keeps its asymptotic behavior, which produces the $AdS_5$ for $y\rightarrow \pm \infty$. A similar non-topological profile was discussed in\cite{nontop}.

\begin{figure}[!ht] 
\centering
\includegraphics[height=4.4cm]{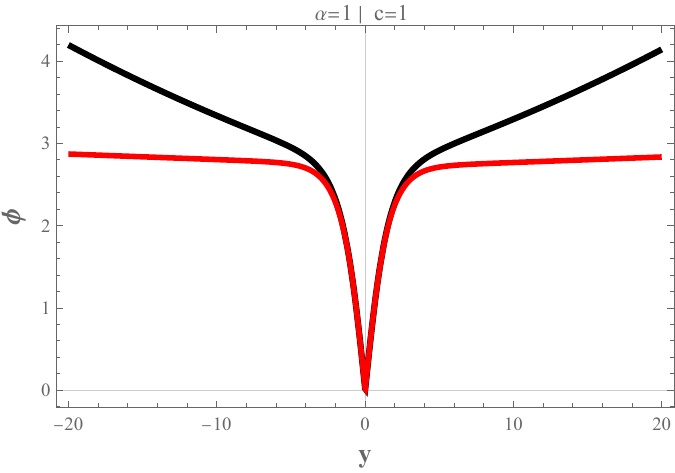}\quad
        \includegraphics[height=4.4cm]{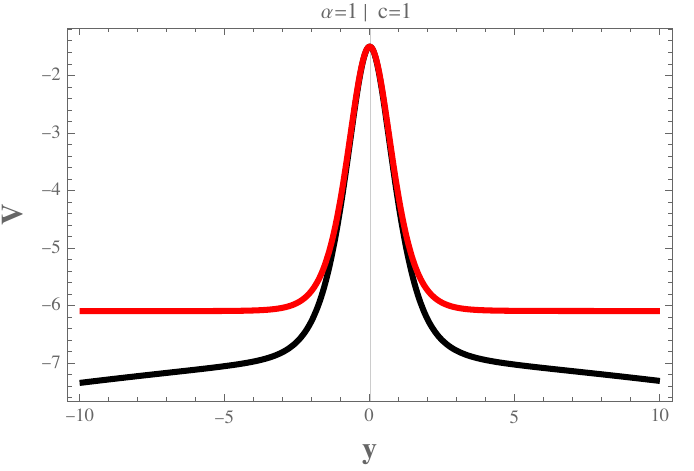}\quad
\caption{Scalar field profile and the potential with respect to $y$ for the scalar Gauss-Bonnet model and the coupling function $\chi=-\frac{M^2}{2}\phi^2$.}  \label{sgb2}
\end{figure}

For the quadratic coupling $\chi=-\frac{M^2}{2}\phi^2$, the Eq.(\ref{thickphiequation}) takes the form
\begin{eqnarray}
\label{phiquadratic}
    (1+24\alpha M^2 A'^2)\phi'^2 &+& 24\alpha M^2 A'^2 \phi \phi'' + 24\alpha M^2 A' (A''-A'^2)\phi \phi' \nonumber\\
    &+& \frac{3}{\kappa}A'' + 12\alpha M^2 A'^2 A'' \phi^2=0.
\end{eqnarray}
The quadratic coupling yields to new non-linear term of form $\phi\phi'$, $\phi\phi''$ and $\phi^2$. We solved the Eq.(\ref{phiquadratic}) for the thick brane anstaz in Eq.(\ref{warpansatz}) for $M^2=10^{-3}$ (black line) and for $M^2=10^{-4}$ (red line) in the fig.(\ref{sgb2}). It turns out that the non-linear terms lead to an asymptotic divergent behavior. Yet, for a small mass-like parameter $M$, the scalar field attains an asymptotic vacuum value, as seen in the red line. Besides, small parameters $M$ also yield to $V(y)<0$, as $y\rightarrow\pm\infty$.  

\subsection{Energy density}

The stress energy tensor for the scalar field is given by
\begin{equation}
    T_{MN}=\partial_M \phi \partial_N \phi -g_{MN}\left(\frac{1}{2}\partial_A \phi \partial^{A}\phi + V(\phi)\right).
\end{equation}
Thus, the non-vanishing components of the stress energy tensor are
\begin{eqnarray}
    T_{00} &=& e^{2A}\left(\frac{1}{2}\phi'^2 + V(\phi)\right)\\
    T_{44} &=& \frac{1}{2}\phi'^2 - V(\phi),
\end{eqnarray}
and $T_{ij}=-T_{00}$. Since the energy density is defined as $\rho=T_{MN}u^{M}u^{N}$, where $u^{M}$ is the unit velocity of a static observer satisfying $g_{MN}u^{M}u^{N}=-1$, by considering $u^{M}=(e^{-A},0,0,0,0)$, the energy density takes the form
\begin{equation}
    \rho=\frac{1}{2}\phi'^2 + V(\phi),
\end{equation}
whereas the pressure along the extra dimension is given by
\begin{equation}
    P_4 =\frac{1}{2}\phi'^2 - V(\phi).
\end{equation}
Along the 3-brane, we have $T_{\mu\nu}=-\rho g_{\mu\nu}$, which is a cosmological constant like equation. In the bulk, the stress energy tensor takes the asymptotic form
\begin{equation}
    T_{MN} \rightarrow - V(\phi_0)g_{MN},
\end{equation}
as $y\rightarrow \pm \infty$. Thus, the bulk is asymptotically $AdS_5$ provided that $V(\phi_0)<0$, as we found before.

\begin{figure}[!ht] 
\centering
        \includegraphics[height=4.4cm]{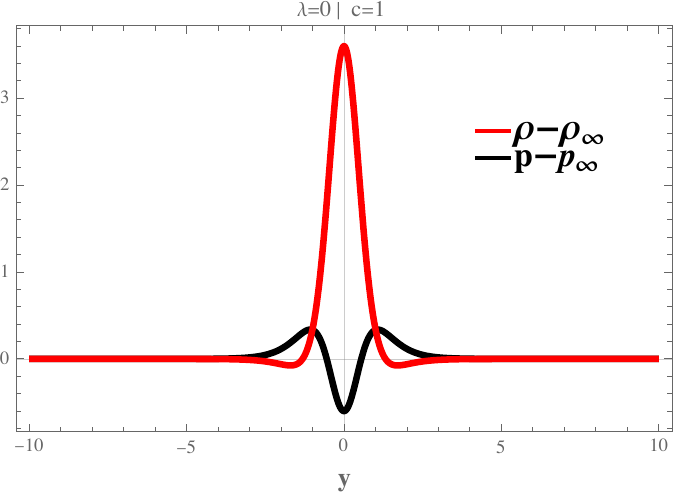}\quad
\caption{Energy and pressue density profiles for the GB model (left) and the difference between the energy and pressure and their vacuum values (right) .}  \label{ep1}
\end{figure}

\begin{figure}[!ht] 
\centering
\includegraphics[height=4.4cm]{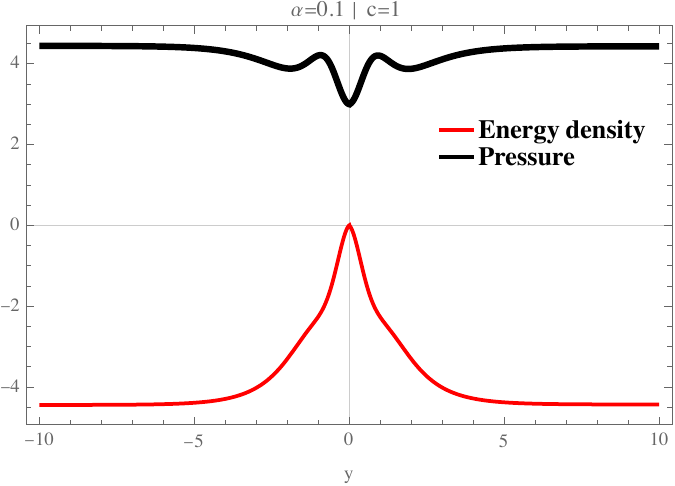}\quad
        \includegraphics[height=4.4cm]{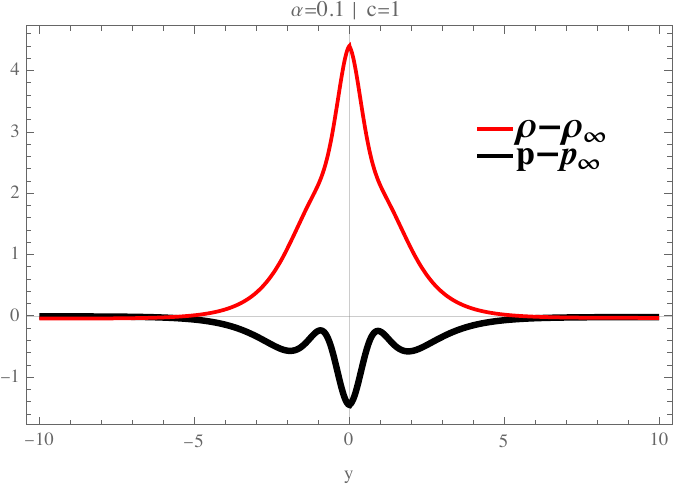}\quad
\caption{Energy density and the pressure profiles for the linear scalar-GB model (left) and their difference (right) .}  \label{ep2}
\end{figure}

\begin{figure}[!ht] 
\centering
\includegraphics[height=4.4cm]{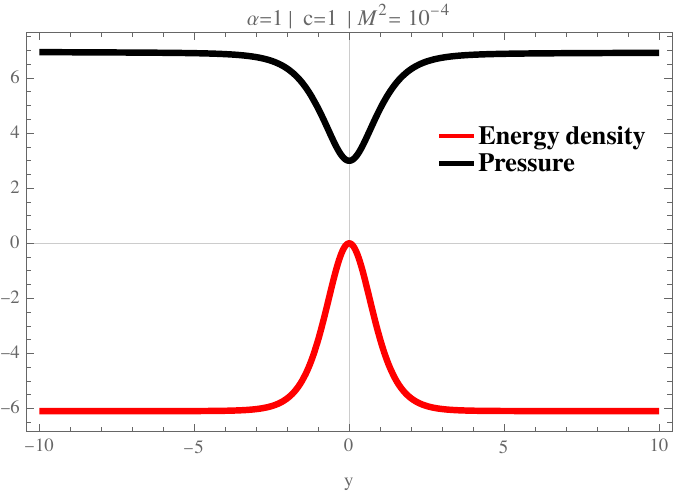}\quad
        \includegraphics[height=4.4cm]{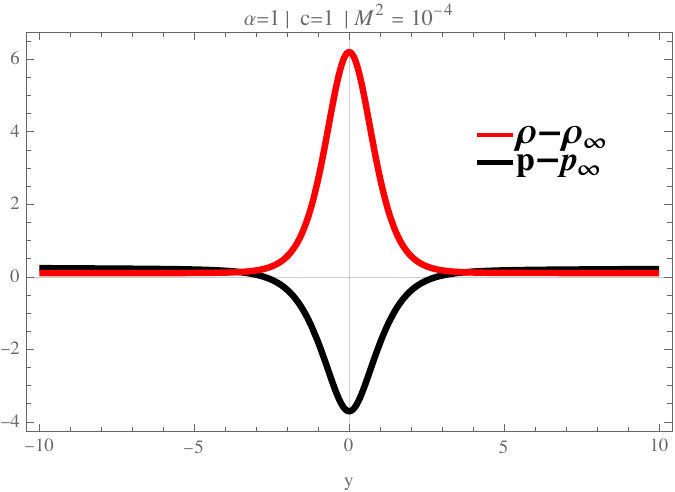}\quad
\caption{Energy density and the pressure profiles for the quadratic scalar-GB model (left). Plot of the difference between the energy and pressure and their vacuum values (right) .}  \label{ep3}
\end{figure}

In order to examine the behaviour of the energy density and the pressure we plotted these quantities for the usual GB model in Fig.(\ref{ep1}), for the linear SGB in Fig.(\ref{ep2}) and for the quadratic SGB in Fig.(\ref{ep3}). In the left panels, we plotted the values of $\rho(y)$ and $P(y)$, whereas in the right panels we plotted the diferences $\rho - \rho_{\infty}$ and $P-P_{\infty}$.

%%%%%%%%%%%%%%%%%%%%%%%%%%%%%%%%%%%%%%%%%%%%%%%%%%%%%%%%%%%%%%%%%%%%%%%%%%%%%%%%%%%
\section{Gravitational perturbations} \label{sec5}

In this section, we investigate the linear stability of the solution against tensor fluctuations of the metric as well as the localization of gravity on the brane. The metric perturbations take the form $\eta_{\mu\nu} \rightarrow \eta_{\mu\nu} + h_{\mu\nu}(x^\mu, y)$, while the scalar field perturbation is expressed as $
\phi \rightarrow \bar{\phi}(y)+\delta\phi(x^\mu, y)$, where $\bar{\phi}$ and $\delta\phi$ stands for the background scalar field and its perturbation, respectively. %In general, perturbations in the tensor, vector, and scalar sectors evolve independently of each other. This allows us to analyze each type separately.
Under tensor perturbations, the metric of the thick brane system takes the form  
\begin{equation}
(ds^{2})_5=e^{2A(y)}\left(\eta_{\mu\nu}+h_{\mu\nu}\right)dx^{\mu}dx^{\nu}+dy^{2}.
\end{equation}
It should be emphasized that, since the theory is diffeomorphic invariant,  we can impose the transverse-traceless (TT) conditions, i.e., $\partial^{\mu}h_{\mu\nu} = \eta^{\mu\nu}h_{\mu\nu} = 0$ \cite{kkgauge}, as performed in other modified gravity models \cite{ft,fq,cubic}. %\red{ which possesses only five degrees of freedom - an appropriate choice for describing a spin-2 particle in $d=5$ . While such gauge fixing is common in $d=5$ theories~\cite{Randall:1999ee,Randall:1999vf}, in $d=6$ scenarios one can additionally encounter scalar and vector modes~\cite{Fu:2018erz}}. 

The nonvanishing components of the perturbed Riemann and Ricci tensors, as well as the Ricci scalar, are given by:
\begin{eqnarray}
\delta R_{\mu\nu\rho\delta}	&=&\frac{1}{2}e^{2A}\left[\partial_{\mu}\partial_{\delta}h_{\nu\rho}-\partial_{\nu}\partial_{\delta}h_{\mu\rho}-\partial_{\rho}\partial_{\mu}h_{\delta\nu}+\partial_{\rho}\partial_{\nu}h_{\delta\mu}\right.\nonumber\\
&+&e^{2A}A'\left(2A'(\eta_{\mu\delta}h_{\nu\rho}-2\eta_{\nu\delta}h_{\mu\rho})+\eta_{\nu\rho}(2h_{\mu\delta}A'+h'_{\mu\delta})\right.\nonumber\\
&-&\left.\left.\eta_{\mu\rho}(2h_{\nu\delta}A'+h'_{\nu\delta})+\eta_{\mu\delta}h'_{\nu\rho}-\eta_{\nu\delta}h'_{\mu\rho}\right)\right],
\end{eqnarray}
\begin{equation}
    \delta R_{\mu4\rho4}=-\frac{1}{2}e^{2A}\left[2h_{\mu\rho}\left(A''+A'^{2}\right)+2A'h'_{\mu\rho}+h''_{\mu\rho}\right],
\end{equation}
\begin{equation}
    \delta R_{\mu\nu\rho4}=\frac{1}{2}e^{2A}\left[\partial_{\mu}h'_{\nu\rho}-\partial_{\nu}h'_{\mu\rho}\right],
\end{equation}
\begin{eqnarray}
    \delta R_{\mu\nu}&=&\frac{1}{2}\left(\partial_{\mu}\partial_{\alpha}h_{\ \!\nu}^{\alpha}+\partial_{\nu}\partial_{\alpha}h_{\ \!\mu}^{\alpha}-\partial_{\mu}\partial_{\nu}h-\Box^{(4)}h_{\mu\nu}\right)\nonumber\\
    &-&\frac{1}{2}e^{2A}\left[h''_{\mu\nu}+A'\left(h'\eta_{\mu\nu}+4h'_{\mu\nu}\right)+2\left(A''+4A'^{2}\right)h_{\mu\nu}\right],
\end{eqnarray}
\begin{equation}
\delta R_{\mu 4}=\frac{1}{2}\left(\partial_{\nu}h'_{\mu}{}^{\nu}-\partial_{\mu}h'\right),
\end{equation}
\begin{equation}
\delta R_{44}=-\frac{1}{2}\left(h''+2A'h'\right),
\end{equation}
\begin{equation}
   \delta R=e^{-2A}\left[\partial_{\mu}\partial_{\nu}h^{\mu\nu}-\Box^{(4)}h\right]+-5A'h'-h'',
\end{equation}where the prime denotes the derivative concerning extra coordinate $y$. Considering the scalar field perturbation, the corresponding perturbation of the energy-momentum tensor is given by:
\begin{eqnarray}
    \delta T_{\mu\nu}&=&-e^{2A}\left(\frac{1}{2}h_{\mu\nu}\bar{\phi}'^{2}+h_{\mu\nu}V+\eta_{\mu\nu}\delta\phi'\bar{\phi}'+\eta_{\mu\nu}\delta\phi V_{\phi}\right),\\
    \delta T_{\mu4}&=&\delta\phi'\partial_{\mu}\bar{\phi}+\bar{\phi}'\partial_{\mu}\delta\phi,\\
    \delta T_{44}&=&\bar{\phi}'\delta\phi'-v_{\phi}\delta\phi.
\end{eqnarray}
Now, let us obtain the equation of motion for the transverse, traceless (TT) tensor perturbation $h_{\mu\nu}$ by employing the perturbed form of the gravitational equation (\ref{EoMmetric}), leading to
\begin{align}\label{kktt}
G(y)h_{\mu\nu}^{\prime\prime}+F(y)h_{\mu\nu}^{\prime}+e^{-2A}H(y)\square h_{\mu\nu}=0, 
\end{align}where
\begin{equation}
    G(y)=1-8\alpha\kappa\left(A'^{2}\chi +2A'\phi'\chi_{\phi}\right),
\end{equation}
\begin{equation}
    F(y)=4A'-8\alpha\kappa\left[2A'\left(2A'^{2}+A''\right)\chi +\left(2A''\phi'+A'\left(9A'\phi'+2\phi''\right)\right)\chi_{\phi}+2\phi'^{2}A'\chi_{\phi\phi}\right],
\end{equation}
\begin{equation}
    H(y)=1-8\alpha\kappa\left[\left(A''+A'^{2}\right)\chi +\left(\phi''+A'\phi'\right)\chi_{\phi}+\phi'^{2}\chi_{\phi\phi}\right].
\end{equation}
Concerning the fluctuations of the scalar field $\phi$, they could in principle be encoded in the factors $G$, $F$, and $H$, since in our model the Gauss–Bonnet term is non-minimally coupled to $\phi$ through the function $\chi(\phi)$. However, we restricted ourselves to analyzing first-order gravitational fluctuations, which exclude second-order contributions in the equation of motion for the perturbation $h_{\mu\nu}$.

It is worthwhile to mention that the non-minimal coupling function $\chi(\phi)$ and its derivatives up to second order modify the Kaluza-Klein equation for the gravitational tensor modes. Moreover, for $\chi=1$, the result obtained in the Ref. \cite{gb} is recovered.

\subsection{Linear Stability} 
After obtained the equation for the tensor modes fluctuation upon the 5 dimensional spacetime, i.e., Eq. (\ref{kktt}), in this section we analyze the propagation of the fluctuations on the 3-brane and along the extra dimension. This analysis is important to ensure that the (3+1) effective action on the 3-brane is well-defined. Moreover, the propagation along the extra dimensions might lead to gravitational instabilities worth to be investigated.

Let us start by performing the usual Kaluza-Klein decomposition, where
\begin{equation}
\label{kkdecomp}
h_{\mu\nu}(x,y)= \chi_{\mu\nu}(x^{\mu})\varphi(y).    
\end{equation}
Substituting Eq.(\ref{kkdecomp}) into the fluctuation equation (\ref{kktt}) yields to
\begin{equation}\label{modos1}
  (\Box^{(4)}  - m^2) \chi_{\mu\nu}(x^\mu)=0
\end{equation}
\begin{equation} \label{modos2}
B^2 \varphi''(y)+C\varphi'(y)+m^2 \varphi(y)=0.
\end{equation}
where
\begin{equation}\label{b}
    B^2 = \frac{e^{2A}G(y)}{H(y)}
\end{equation}
and 
\begin{equation}\label{c}
    C = \frac{e^{2A}F(y)}{H(y)}.
\end{equation}
The Eq. (\ref{modos1}) describes the (3+1) KK graviton with mass $m$ on the 3-brane, whereas the Eq.(\ref{modos2}) governs the graviton propagation along the extra dimension. In order to keep the causality behavior of the KK modes governed by the Eq.(\ref{modos2}), we impose the condition that $G(y)>0$. Hence, we assume the constrain
\begin{equation}
    \chi \left(A'^2 + 2 A' \phi' \frac{\chi_\phi}{\chi} \right) < \frac{1}{8\kappa \alpha}, 
\end{equation}
what yields an upper bound to the non-minimal coupling. Asymptotically, for $A'=-c$ leads to $    \chi \left(1 - 2 \frac{\phi'}{c} \frac{\chi_\phi}{\chi} \right) < \frac{1}{8\kappa \alpha c^2}$. For the usual EGB, the upper bound is $\alpha <\frac{1}{8\kappa c^2}$

The propagation of the tensor fluctuations along the 3-brane in Eq.(\ref{modos1}) has a mass term steaming from the extra dimension. This mass, known as the KK mass, can be obtained using the propagation along the extra dimension Eq.(\ref{modos2}). In order to analyze the stability of the KK spectrum, we employ the analogue supersymmetric quantum mechanics approach.

Let us perform the change of coordinates
\begin{equation}
\label{coordinate}
z=\int B^{-1}(y) dy,     
\end{equation}
where $B(y)$ is given by Eq.(\ref{b}). As a result, 
the Eq.(\ref{modos2}) assumes the form
%To demonstrate that Eq. (\ref{modos2}) describes non-tachyonic KK modes, i.e., stable modes, we assume the following coordinate transformation $dy=B(y(z))dz$. At this juncture in the text, it is pivotal to underscore a crucial observation concerning the coordinate transformation between the variables $y$ and $z$. The functional correlation between these variables is contingent upon the function $B$, which, in turn, is intricately linked to the warped factor $A$. Put simply, the existence of a bijective relationship hinges on the specific choice of $A$, thereby influencing the smoothness of the transition between $y$ and $z$. Owing to this constraint, our examination of KK modes will involve their analysis in the variable $y$ utilizing Eq. (\ref{modos2}). %Notably, for the particular case under consideration in this paper—specifically, the solutions derived from the superpotential $W=\xi \phi$ and for solution (\ref{sec})— the Fig. (\ref{b})  demonstrates a bijective transition between coordinates, validating the smoothness of the transformation.  
%After substituting this change of coordinates into Eq.(\ref{modos2}), we find that 
%\begin{figure}[!ht] 
%\includegraphics[height=5cm]{yz.pdf}\quad
%\caption{Coordinate $z(y)$ for $\epsilon=0.5$ considering different trick brane solutions.}  \label{b}
%\end{figure}
\begin{equation}
    \ddot{\varphi}(z) + E(z)  \dot{\varphi}(z)+m^2 \varphi(z)=0,
\end{equation}
where $E(z)= \frac{C - \dot{B}}{B}$ and the dot represents the derivative with respect to the $z$ coordinate.  Now assuming the following change on the wave function 
\begin{equation}
 \varphi(z) = e^{-\frac{1}{2}\int Edz} \Phi(z),    
\end{equation}
then the function $\Phi$ satisfies a Schr\"{o}dinger-like equation given by \cite{kk}
\begin{equation} \label{susy}
    (-\partial_{z}^2+ V_{eff})\Phi = m^2\Phi,
\end{equation}
where the effective potential $V_{eff}$ is given by
 \begin{equation}
     V_{eff}(z) = \bigg( \frac{E}{2} \bigg)^2 + \partial_{z} \bigg( \frac{E}{2} \bigg).
 \end{equation}
The form of the effective potential shows that it possesses a supersymmetric analogue structure of quantum mechanics. This SUSY-like symmetry guarantees the stability analysis for the KK modes \cite{kk}. 
Indeed, by defining a superpotential $U_g = -\frac{E}{2}$ and the SUSY-like Hamiltonian operator $Q^{\dagger} Q$, where
$ Q = \partial_{z} + U_g $, then the Schr\"{o}dinger-like equation Eq.(\ref{susy}) takes the form
\begin{equation} \label{eqq}
    Q^{\dagger} Q \Phi = m^2 \Phi. 
\end{equation}
Therefore, the mass eigenvalue is positive, $m^2\ge 0$, and thus, there is no tachyonic KK gravitational modes. Accordingly, we conclude that the cubic modifications preserves the graviton stability, at least at the linearized regime.

%In other words, we demonstrate that the cubic gravity in the scenario of braneworld presented by us is stable in the linear regime. Note that it is straightforward to incorporate the effects of GB gravity into this analysis. Given the similarity between the forms of Eq. (\ref{modosgb}) and Eq. (\ref{modos}), the alteration due to this addition primarily manifest as changes in the functions $B^2$ and $C$. Consequently, in a more general context, we establish the stability even in the presence of cubic interaction. 

\subsection{Massive and massless modes}
Once we determined the stability of the KK spectrum, 
now we explore the effects of Einstein-Scalar-Gauss-Bonnet gravity on these massive and massless KK modes. Unfortunately, the change of coordinate $z=z(y)$ in Eq.(\ref{coordinate}) cannot be obtained analytically for the thick brane solutions. Thus, we study the modes in the $y$ coordinate, using the Eq.(\ref{modos2}).

We begin by computing the massive modes. Let us consider the thick brane scenario generated by a warped factor~\eqref{warpansatz}. By substituting this warp factor into Eqs.~\eqref{b} and~\eqref{c}, we obtain:
\begin{equation} \label{B}
    B^2 = \frac{\text{sech}^2(c y) \left(1-8 \alpha  \kappa  \left(c^2 \chi  \tanh ^2(c y)-2 c \chi _{\phi } \phi ' \tanh (c y)\right)\right)}{1-8 \alpha  \kappa  \left(\chi  \left(c^2 \tanh ^2(c y)-c^2 \text{sech}^2(c y)\right)+\chi _{\phi } \left(\phi ''-c \phi ' \tanh (c y)\right)+\left(\phi '\right)^2 \chi _{\phi \phi }\right)},
\end{equation}
\begin{align} \label{C}
        &C = \frac{4 c \tanh (c y) \text{sech}^2(c y) \left(4 \alpha  \kappa  \left(\phi '\right)^2 \chi _{\phi \phi }+8 \alpha  c^2 \kappa  \chi -12 \alpha  c^2 \kappa  \chi  \text{sech}^2(c y)-1\right)}{-8 \alpha  c^2 \kappa  \chi +16 \alpha  c^2 \kappa  \chi  \text{sech}^2(c y)-8 \alpha  \kappa  \left(\chi _{\phi } \left(\phi ''-c \phi ' \tanh (c y)\right)+\left(\phi '\right)^2 \chi _{\phi \phi }\right)+1} \\ \nonumber
        & + \frac{8 \alpha  c \kappa  \chi _{\phi } \text{sech}^2(c y) \left(2 \phi '' \tanh (c y)+c \phi ' \left(11 \text{sech}^2(c y)-9\right)\right)}{-8 \alpha  c^2 \kappa  \chi +16 \alpha  c^2 \kappa  \chi  \text{sech}^2(c y)-8 \alpha  \kappa  \left(\chi _{\phi } \left(\phi ''-c \phi ' \tanh (c y)\right)+\left(\phi '\right)^2 \chi _{\phi \phi }\right)+1}
\end{align}

\begin{figure}[h] 
\centering
       \includegraphics[height=4.9cm]{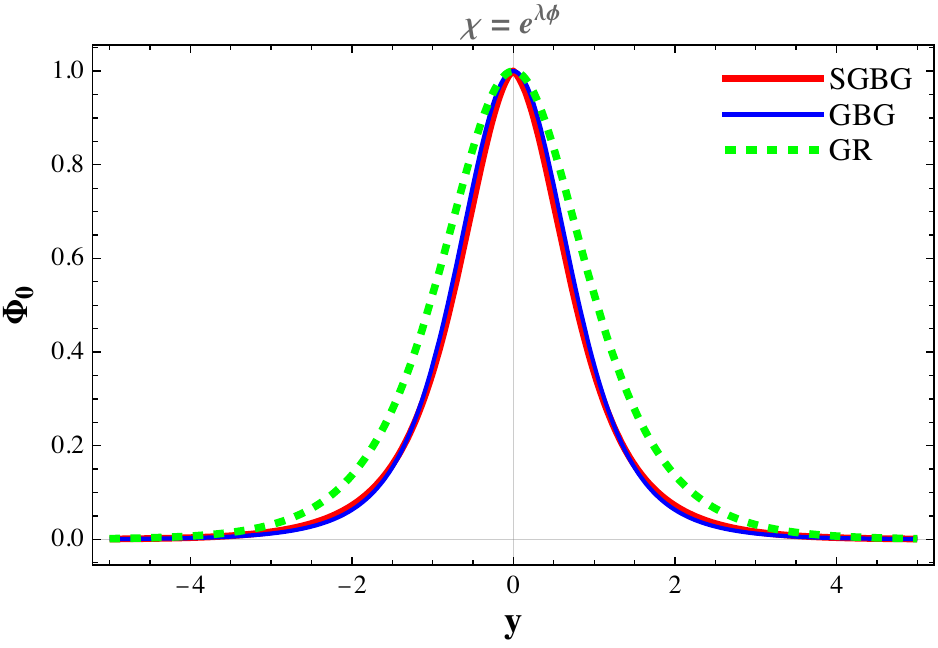}\quad
        \includegraphics[height=4.9cm]{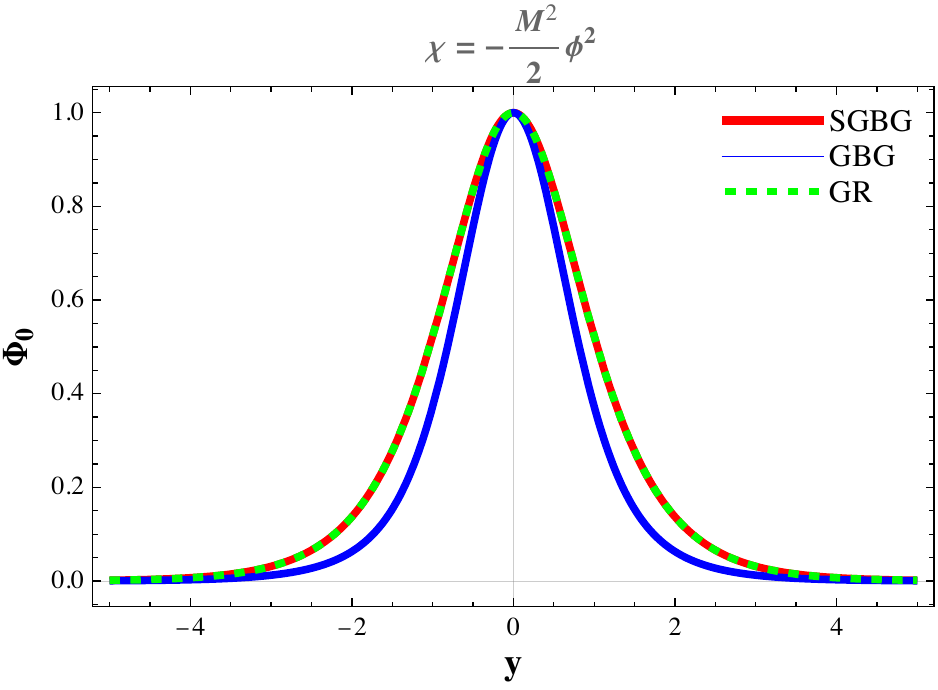}\quad
           \caption{Massless KK modes generated by a thick brane. We assume that  $c=1$,$\alpha=0.1$, $\lambda=0.1$ and $M^2=10^{-4}$.} 
          \label{massless}
\end{figure}

Finally, let us consider the massless mode $m=0$. This mode must be confined to the brane to ensure that our theory leads to a finite action in the dimensions $(3+1)$, that is, this zero mode is responsible for reproducing the massless gravity in $d=4$ \cite{cubic,kk}. From Eq. (\ref{eqq}), the massless mode can be expressed as:
\begin{equation}
    \Phi_{0}(y) = \Tilde{\Phi}_0 e^{- \int U_{g}dz} 
\end{equation}
where $\Tilde{\Phi}_0$ is a normalization constant. Further, the localized graviton zero mode should satisfy the condition $\int_{-\infty} ^{-\infty} \Phi_0 ^2 dz=1$. Now, employing the $\frac{dy}{dz}= B(y)$ transformation, we proceed to numerically generate graphs depicting the zero modes as functions of the variable $y$. It is essential to highlight that the corresponding solution in this variable is expressed as follows

\begin{equation}
    \Phi_{0}(z) = \Tilde{\Phi}_0 e^{- \int U_{g}(y)dy}, 
\end{equation}
where
\begin{equation}
    U_{g}(y) = - \frac{C - B'B}{2 B^2}.
\end{equation}
Indeed, by substituting the Eqs.~\eqref{B} and~\eqref{C} into the above equation, we numerically obtain the plots (\ref{massless}), which demonstrate that the massless modes remain localized on the brane, just as in General Relativity \cite{kk}. This result confirms that the non-trivial introduction of the scalar field coupled to the Gauss-Bonnet curvature term still preserves field localization on the brane.

\begin{figure}[h] 
\centering
       \includegraphics[height=4.9cm]{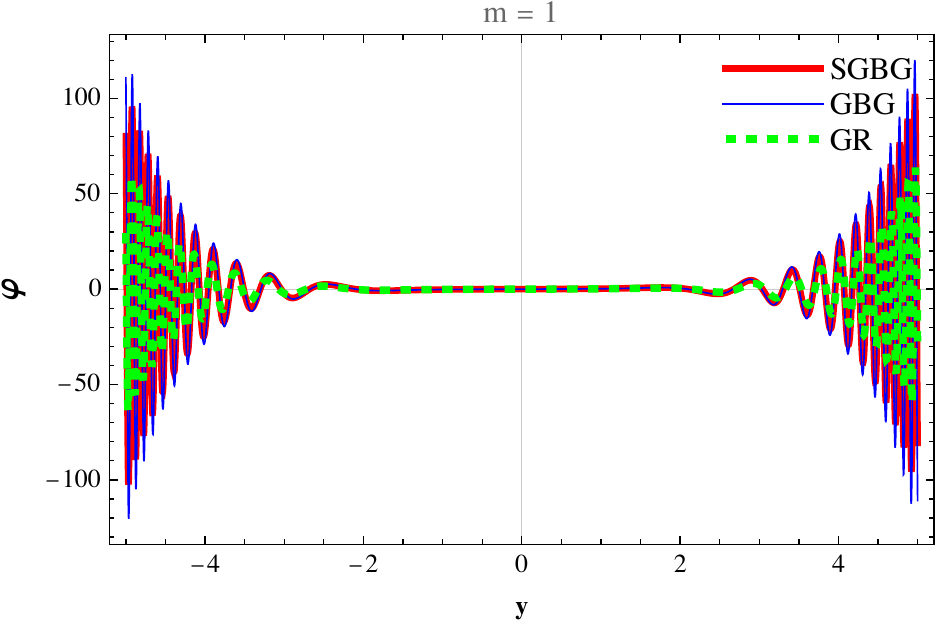}\quad
        \includegraphics[height=4.9cm]{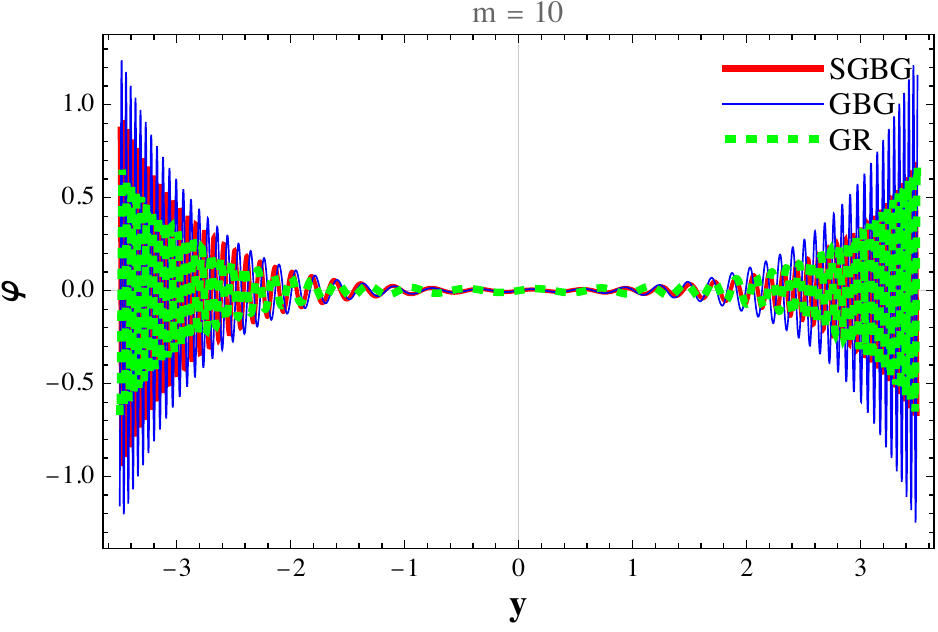}\quad
           \caption{Massive KK modes generated by a thick brane with dilaton-like coupling (\ref{dilatonchi}). We assume that  $c=1$, $\alpha=0.1$ and $\lambda=0.1$.} 
          \label{mass}
\end{figure}
\begin{figure}[h] 
\centering
       \includegraphics[height=4.9cm]{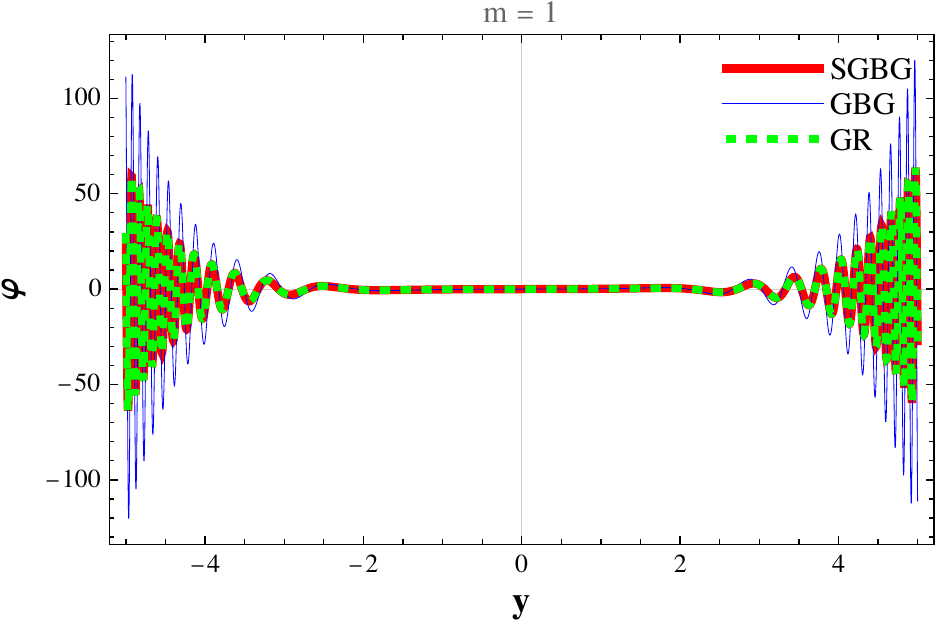}\quad
        \includegraphics[height=4.9cm]{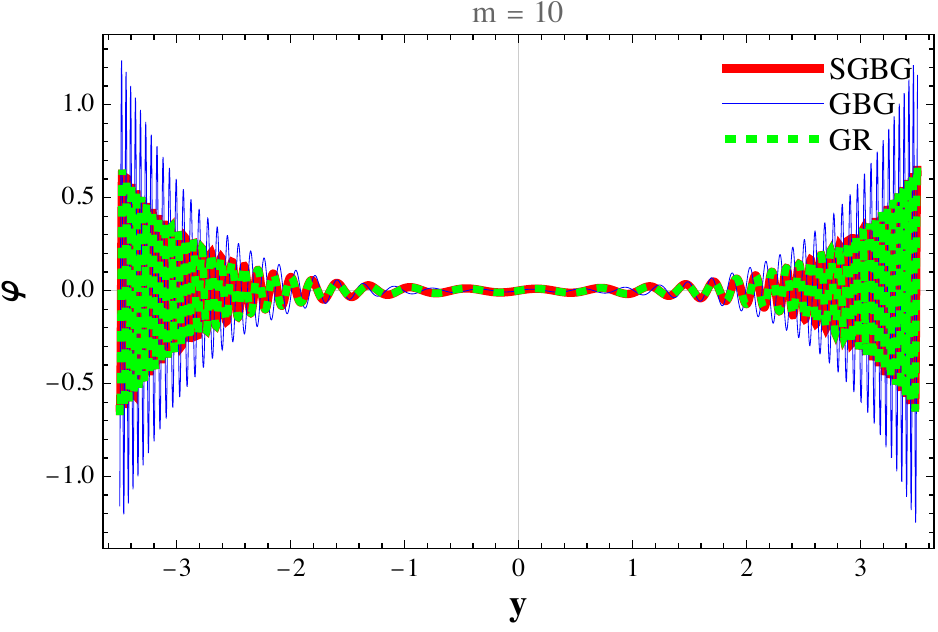}\quad
           \caption{Massive KK modes generated by a thick brane with quadratic coupling (\ref{pdm}). We assume that  $c=1$, $\alpha=0.1$, $\lambda=0.1$ and $M^2=10^{-4}$ .} 
          \label{mass1}
\end{figure}

\section{Final remarks}  \label{con}
In this work, we explore the theoretical consequences of Einstein-Scalar-Gauss-Bonnet (ESGB) gravity in braneworld models. Our analysis reveals that the gravitational dynamics of warped braneworlds are profoundly modified by the non-minimal coupling $\chi(\phi)$ between a scalar field and the Gauss-Bonnet curvature term. Specifically, the choice of coupling function critically determines the scalar field’s role in defining the thick brane’s physical properties.

We found that outside the brane core, where the scalar field reaches a constant state $\phi_0$, the spacetime geometry has a asymptotic $AdS_5$ behavior, where the bulk cosmological constant $\Lambda =2\kappa V(\phi_0)$ is dynamically generated. Moreover, even for $\Lambda=0$, the introduction of the scalar-curvature coupling leads to a non-trivial asymptotic RS solution. A similar behavior was found in other modified gravity theories, as in cubic corrections \cite{cubic}.

We examined thick brane properties by considering the scalar field dynamics. The non-minimal coupling includes new terms in the modified gravitational and scalar field equations compared to the usual EGB braneworld \cite{gb}. As a result, the braneworld solutions are quite sensitive to $\chi(\phi)$ choice. We considered two possible couplings: one linear and another quadratic in the scalar field. For both couplings, we adopted the thick brane warp factor ansatz~\eqref{warpansatz} and numerically solved the scalar field equation. The  $\chi(\phi)$ coupling function modifies the asymptotic behavior of the scalar field. As a result, the scalar field does not exhibit a domain-wall pattern. For the linear coupling~\eqref{dilatonchiapprox}, the scalar field has a parity-even profile converging to a constant value asymptotically. For the quadratic coupling both the scalar field and its potential converge to a constant vacuum state, provided that the coupling be small. Moreover, the energy and pressure densities have an asymptotic $AdS_5$ behavior, showing that the cosmological constant is dynamically generated by the scalar field. By subtracting their asymptotic value we obtain the brane core energy and pressure densities, which exhibit a localized profile. 

Finally, we investigate the stability of the model by analyzing the dynamics of linear gravitational perturbations. After deriving the linearized gravitational field equations, we observe that their dynamics are linked to the scalar-curvature coupling constant. Furthermore, by employing a supersymmetric-like method from quantum mechanics, we demonstrated the stability of our model. This implies the absence of tachyonic Kaluza-Klein (KK) modes in our proposed framework.

Using the equations for tensor fluctuation modes, we studied their localization on the brane. In this case, we assumed a thick brane as described in the ansatz~\eqref{warpansatz}. As expected, the modes are significantly modified, even in the linear regime, by the choice of the $\chi(\phi)$ form. Interestingly, the massive modes (e.g., for exponential coupling) exhibit parity-odd Kaluza-Klein (KK) states. On the other hand, the localization of massless modes closely resembles that of standard Gauss-Bonnet theory.
 An interesting development of this work is a gauge-independent analysis of the gravitational perturbation, considering the scalar and the vector modes, as performed in \cite{kkgauge}.

%%%%%%%%%%%%%%%%%%%%%%%%%%%%%%%%%%%%%%%%%%%%%%%%%%%%%%%%%%%%%%%%%%%%%%%%%%%
\section*{Acknowledgments}
\hspace{0.5cm} The authors thank the Funda\c{c}\~{a}o Cearense de Apoio ao Desenvolvimento Cient\'{i}fico e Tecnol\'{o}gico (FUNCAP), Fundação de Amparo à Pesquisa e ao Desenvolvimento Científico e Tecnológico do Maranhão (FAPEMA), the Coordena\c{c}\~{a}o de Aperfei\c{c}oamento de Pessoal de N\'{i}vel Superior (CAPES), and the Conselho Nacional de Desenvolvimento Cient\'{i}fico e Tecnol\'{o}gico (CNPq), Grant no. 311393/2025-0 (RVM).  L.A.L is supported by FAPEMA BPD- 08975/24.

%%%%%%%%%%%%%%%%%%%%%%%%%%%%%%%%%%%%%5%%%%%%%%%%%%%%%%%%%%%%%%%%%%%%%%%%%5      END OF THE MANUSCRIPT 
%%%%%%%%%%%%%%%%%%%%%%%%%%%%%%%%%%%%%%%%%%%%%%%%%%%%%%%%%%%%%
%\cite{Penrose:1964wq}
%\cite{Berti:2015itd}

\end{document}